\documentclass[a4paper]{jpconf}
\usepackage{graphicx}
\begin{document}
\title{Geophysical constraint on a relic background of the dilatons}

\author{Sachie Shiomi\footnote{Present address: Institute for Cosmic Ray Research, The University of Tokyo, Kashiwa, Chiba 277-8582, Japan.}}

\address{Space Geodesy Laboratory, Department of Civil
Engineering, National Chiao Tung University, Hsinchu, Taiwan 300,
Republic of China}


\begin{abstract}
According to a scenario in string cosmology, a relic background of light dilatons can be a significant component of the dark matter in the Universe. 
A new approach of searching for such a dilatonic background by observing Earth's surface gravity was proposed in my previous work. 
In this paper, the concept of the geophysical search is briefly reviewed, and the geophysical constraint on the dilaton background is presented as a function of the strength of the dilaton coupling, $q_b^2$. 
For simplicity, I focus on massless dilatons and assume a simple Earth model. With the current upper limit on $q_b^2$, we obtain the upper limit on the dimensionless energy density of the massless background, $\Omega_{DW}h^2_{100} \leq 6 \times 10^{-7}$, which is about one-order of magnitude more stringent than the one from astrophysical observations, at the frequency of $\sim$ 7 $\times$ 10$^{-5}$ Hz. If the magnitude of $q_b^2$ is experimentally found to be smaller than the current upper limit by one order of magnitude, the geophysical upper limit on $\Omega_{DW}h^2_{100}$ becomes less stringent and comparable to the one obtained from the astrophysical observations.
\end{abstract}

\section{Introduction}

The scalar gravitational fields (or the dilatons) have appeared in
scalar-tensor theories of gravitation and also in theories towards
the unification of the fundamental forces in nature. These theories 
predict the existence of scalar waves. Their sources fall into two categories. One is
cosmological and the other is astronomical. The former is predicted
as a relic background of the dilatons in string cosmology
\cite{Gasperini2003}. The latter is as emission from a spherical
symmetric gravitational collapse of a star \cite{Shibata1994,
Harada1997}. The detection or experimental constraints on the scalar waves 
would play an important role for the development of the unified theories and the verification of gravitational theories.
In this paper, I focus on the search for scalar waves from the cosmological origin. 

Scalar waves can interact with detectors in two ways
\cite{Gasperini1999}: (1) indirectly, through the geodesic coupling
of the detectors to the scalar component of the metric fluctuations (e.g. \cite{Maggiore2000})
and (2) directly, through the effective dilatonic charges of the
detectors, which depend on the internal chemical composition of the
detectors. In the direct coupling, the response of the detectors is
nongeodesic in general \cite{Gasperini1999}. A new approach of searching for the direct-coupling scalar waves using the geophysical
tool of superconducting gravimeters has been proposed 
\cite{Shiomi2008}. Here, I briefly review the concept
of the geophysical search (Section \ref{Concept}). Also, I give a further discussion on the preliminary upper limits on the dilaton background, obtained in my previous work \cite{Shiomi2008}, 
and present a geophysical constraint on a relic background of massless dilatons (Section \ref{Geophysical constraint}).

\section{Geophysical search for composition-dependent dilatonic waves}\label{Concept}

\subsection{The concept}

\begin{figure}
\includegraphics[width=\linewidth]{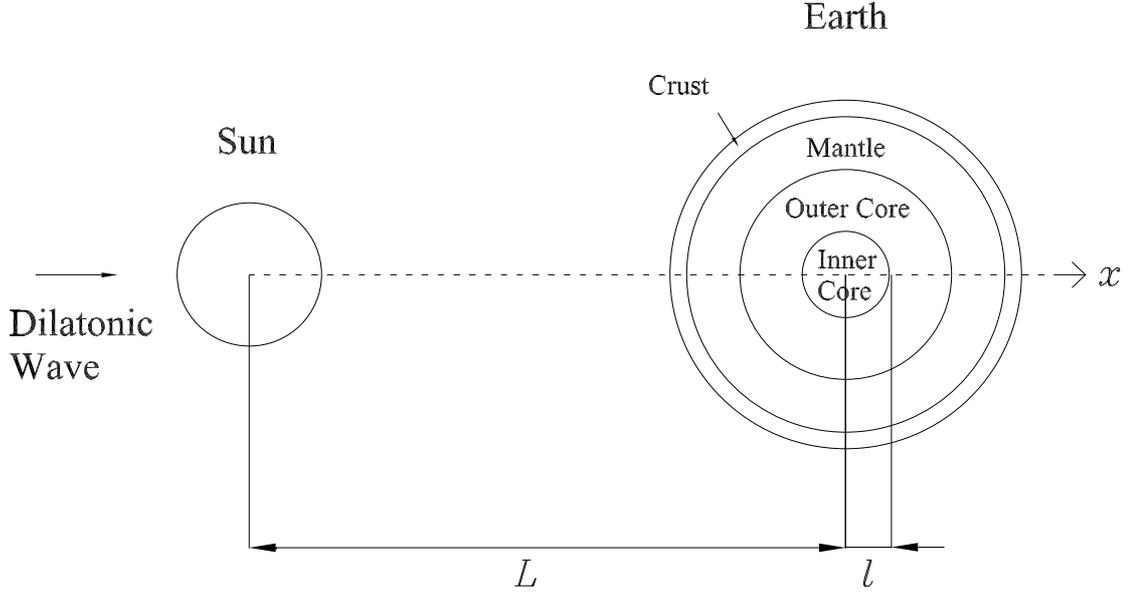}\caption{The concept of the geophysical search for dilatonic waves (not drawn to scale), 
quoted from \cite{Shiomi2008}. 
As dilatonic waves propagate along the Sun-Earth line,
the Earth's inner core oscillates with amplitude $l$. Such oscillations would cause changes in surface gravity, which can be searched for with superconducting gravimeters installed at surface of the Earth.}
\label{fig:concept}
\end{figure}

The Earth's interior can mainly be classified into four layers: the
crust, the mantle, the liquid outer core, and the solid inner core (see Figure \ref{fig:concept}).
The solid inner core is held at the center of the Earth mainly by
the gravitational pull from the rest part of the Earth. 

Unified theories predict the existence of dilatonic charges that depend on the internal chemical structure of matter. 
It should be noted that the dilatonic charge of the inner core is
different from the one of the rest part of the Earth, because of
the difference in their chemical compositions (see Equation (\ref{eq:q}) below for the definition of dilatonic charges). 
The inner core is made from dense elements, such as Iron and Nickel and its average density is
13,000 kg m$^{-3}$, while the rest part of the Earth is mainly made
from less dense materials, such as silicon oxides, and its average
density is about 5,400 kg m$^{-3}$. 

Because of the difference in the dilatonic charges, impinging dilatonic waves on the Earth would couple differently to the inner core and the rest part of the Earth. 
As a result, there would be relative translational motions between the inner core and the rest part of the Earth, or oscillations of the inner core relative to the rest part of the Earth.
Such relative motions cause surface gravity changes that can be searched for with superconducting gravimeters, the most sensitive instruments to measure surface gravity at the frequencies of our interest. 

Unlike the tensor gravitational waves predicted by general relativity, which are transverse, dilatonic
waves are longitudinal waves. When dilatonic waves propagate along the Sun-Earth line, we would observe strains due to the dilatonic waves along the line (Figure \ref{fig:concept}). 
The maximum amplitude of the translational motion would be observed when the dilatonic waves have the same frequencies as the natural oscillation frequencies of the inner core. 

\subsection{Equation of motion}\label{st:Equation of motion}

The equation of motion of the inner core can be given by (Equation (1) of
\cite{Shiomi2008}):
\begin{equation}
\ddot{l^i} \approx -\gamma \dot{l^i} - \omega_0^2 l^i - \Delta q
(L^k \partial_k + 1)\partial^i \phi, \label{eq:EofM}
\end{equation}
where $i$ and $k$ indicate the space-time components of the
parameters, $\Delta q$ is the difference in dilatonic charge between
the inner core and the rest part of the Earth (for the definition,
see Equation (\ref{eq:delta_q}) below.), $L$ is the rest separation between the Sun and
the Earth (see Figure \ref{fig:concept}), and $\phi$ is the dilaton
field. The gravitational stiffness and the friction
coefficient are respectively given by
\begin{eqnarray} \omega_0^2 &\approx& \frac{4}{3} \pi G
\frac{\rho_{ic} - \rho_{oc}}{\rho_{ic}} \rho_{oc} \approx 1.8 \times
10^{-7} \rm{s^{-2}} \approx \{2 \pi (4.1 \hspace{5
pt}\rm{h})^{-1}\}^2, \label{eq:omega_0^2}\\
\gamma &\equiv& \frac{6 \pi \eta r_{ic}}{m_{ic}} \approx 2.3 \times
10^{-16} \eta \hspace{5 pt} \rm{s^{-1}},
\end{eqnarray}
where $G = 6.67 \times 10^{-11}$ N m$^2$ kg$^{-2}$ is the
gravitational constant, $m_{ic}$ $\approx$ 9.8 $\times$ 10$^{22}$ kg
is the mass of the inner core, and $\eta$ is the effective viscosity
of the outer core.

Dilatonic charges can be defined as a relative strength of scalar to
gravitational forces, by summing over all the components $n$ of the
test body (Equation (22) of \cite{Gasperini1999}):
\begin{equation}
q = \frac{\sum_n m_n q_n}{\sum_n m_n}.\label{eq:q}
\end{equation}
For a test body $A$ made from an ordinary matter with mass $M$, the
dilatonic charge is given by
\begin{equation}
q_A \simeq \frac{B m_b q_b}{M} \equiv \frac{B}{\mu} q_b,
\end{equation}
where $B$, $m_b$, and $q_b$ are the baryon number, baryon mass, and
dimensionless-fundamental barionic charge, respectively. The
difference in dilatonic charge between the inner core ($ic$) and the
rest part of the Earth ($rpe$) can be given by
\begin{equation}
\triangle q \simeq \left\{\left(\frac{B}{\mu}\right)_{ic} -
\left(\frac{B}{\mu}\right)_{rpe}\right\}q_b \equiv \triangle
\left(\frac{B}{\mu}\right)q_b.\label{eq:delta_q}
\end{equation}
The magnitude of $q_b$ is constrained by tests of the equivalence principle \cite{Schlamminger2008}.

\subsection{Estimation of the upper limits}

The spectrum of the displacement of the inner core ($l$) at
resonance can be given by (Equation (6) of \cite{Shiomi2008}):
\begin{equation}
S_l (f_0) = \frac{(\Delta q)^2 (L^2 \kappa^2 +1)\kappa^2
c^4}{4\omega_0^2 \gamma^2} S_h(f_0) \label{eq:S_l}
\end{equation}
where $c$ is the speed of light and $\kappa = \omega_0/c$ is the
wave length. The resonant frequency is $f_0 = \omega_0/2 \pi \simeq
6.8 \times 10^{-5}$ Hz. $S_h(f)$ is the power spectrum of the strain
due to massless dilatonic waves: $h \equiv 2 \phi/c^2$.

From residual gravity data of superconducting gravimeters, an upper
limit on the spectrum of the displacement is given to be $S_l (f_0)
< 1.1 \times 10^{-4}$ mHz$^{-1/2}$. The dimensionless energy density for a stochastic background of
massless dilaton is related to the strain spectrum by
\begin{equation}
\Omega_{DW}(f) = \frac{\pi^2 f^3}{3 H_0^2} S_h(f)\label{eq:Omega_DW}
\end{equation}

With the equation (\ref{eq:S_l}) and the upper bound on the
displacement from gravity data, we obtain the upper limit on the
dimensionless energy density \cite{Shiomi2008}:
\begin{equation}
\Omega_{DW}(f_0)h_{100}^2 < 2.2 \times 10^{-17} \eta^2.
\label{eq:Omega_DW_limit}
\end{equation}
This upper limit is proportional to $\eta^2$. However, the magnitude
of $\eta$ is not well determined. A handbook of physical
constants \cite{Secco1995} shows different values of $\eta$, ranging
from 10$^{-3}$ Pa s to 10$^{12}$ Pa s, depending on different
methods to estimate the magnitude. A more resent estimate of an
upper bound on $\eta$ given by nutation data is $\eta < 1.7 \times
10^5$ Pa s \cite{Mathews2005}.

Using this resent upper bound from nutation data, we obtain
$\Omega_{DW}h^2_{100} \leq 6 \times 10^{-7}$ at the resonant
frequency. This is below the astrophysical constraints on a massless background obtained from
Nucleosynthesis and measurements of the cosmic microwave background
\cite{Cyburt2005}: $\Omega h^2_{100} \leq 10^{-5}$. 
Expected upper limits on the massless dilaton background for different values of the effective viscosity is given in
Figure 2 of \cite{Shiomi2008}.

\section{Geophysical constraint on the massless dilaton background}\label{Geophysical constraint}

\begin{figure}
\includegraphics[width=\linewidth]{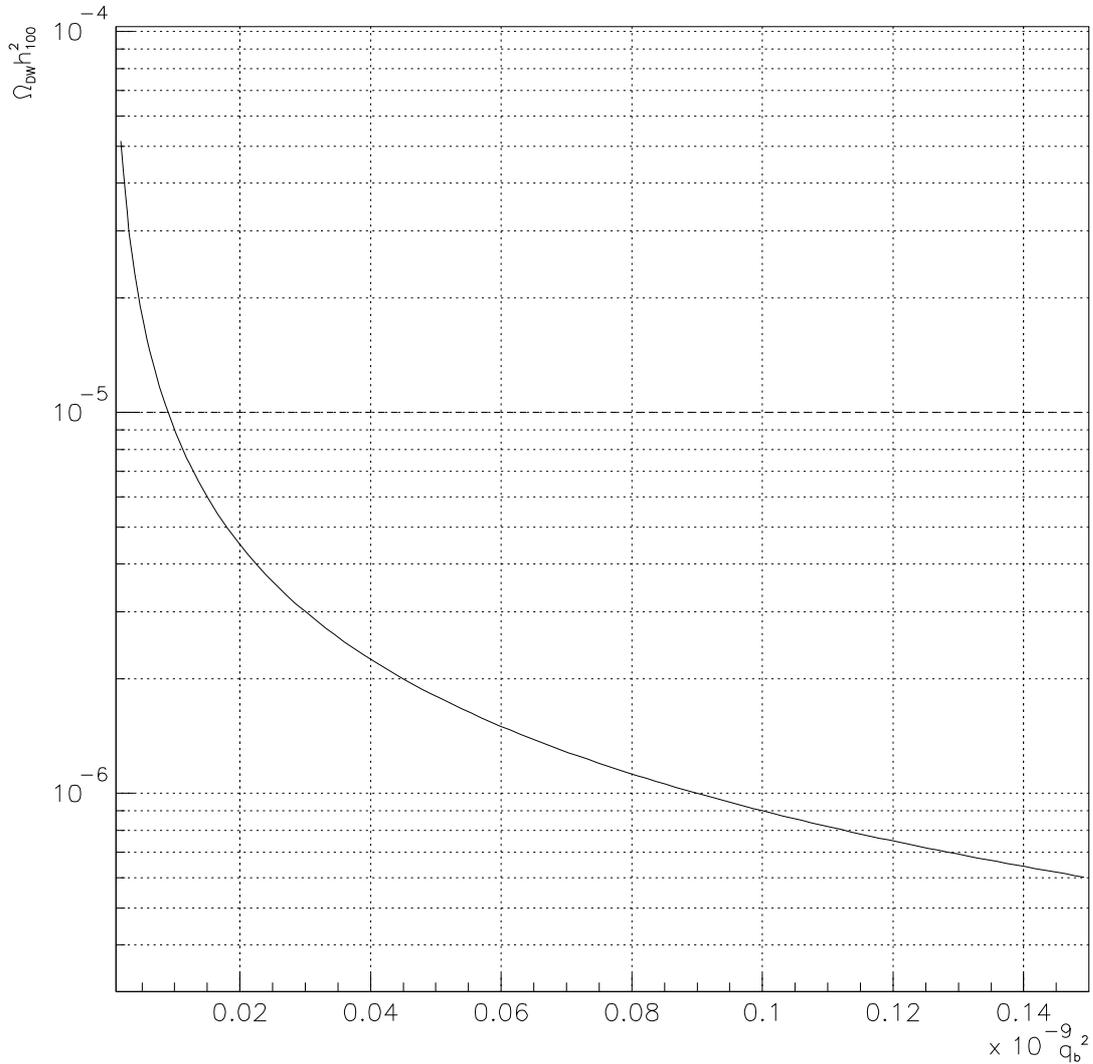}\caption{Constraint on the dimensionless energy density of the
massless dilaton background (the solid curve) at the resonant frequency of $\sim$ 7 $\times$ 10$^{-5}$ Hz, obtained using the recent upper bound
on the effective viscosity: $\eta \simeq$ 1.7$\times$10$^5$ Pa s
(estimated from nutation data \cite{Mathews2005}). The dotted line indicates the astrophysical upper limit on the massless background \cite{Cyburt2005}.
The region below the dotted line and above the solid curve is constrained by this geophysical search. 
The strength of the dilaton coupling is constrained to be $q_b^2 \leq 1. 5 \times 10^{-10}$ by the tests of the equivalence principle \cite{Schlamminger2008}.}
\label{fig:Constraint}
\end{figure}

From the equation (\ref{eq:S_l}), one can see that the upper limit on the dimensionless energy density depends on the magnitude of the strength of the dilaton coupling $q_b^2$. 
In the above estimation of the upper limit on the dimensionless energy density (the condition (\ref{eq:Omega_DW_limit})), we have used the upper bound of $q_b^2 \sim 1.5 \times 10^{-10}$ at the Yukawa range of our interest, obtained by tests of the equivalence principle \cite{Schlamminger2008}. In Figure \ref{fig:Constraint}, the upper limit on the massless dilaton background at the resonant frequency of $\sim$ 7 $\times$ 10$^{-5}$ Hz is drawn against the strength of the dilaton coupling $q_b^2$, using the recent estimate of the upper bound on the effective
viscosity of $\eta \sim 1.7 \times 10^5$ Pa s.
%
%
At the frequency of our interest, there are the astrophysical constraints on the massless background (the dotted line in Figure \ref{fig:Constraint}), obtained from Nucleosynthesis and measurements of cosmic microwave background \cite{Cyburt2005}. The astrophysical constraints are independent of the strength of the dilaton coupling. The region below the dotted line and above the solid curve is constrained by the geophysical search.

\section{Discussions and Summary}

In my previous work \cite{Shiomi2008}, I have proposed the geophysical search for dilatonic waves and presented preliminary upper limits on a massless dilatonic background by assuming a simple Earth model. In this work, I have discussed the upper limit further and presented the constraint on the dimensionless energy density of the massless dilaton background as a function of the strength of the dilaton coupling, $q_b^2$ (Figure \ref{fig:Constraint}).

The magnitude of $q_b^2$ is constrained by tests of the equivalence principle \cite{Schlamminger2008}. By using the most stringent constraint on $q_b^2$, I have obtained the upper limit on the dimensionless energy density of a massless dilaton background: $\Omega_{DW}h^2_{100} \leq 6 \times 10^{-7}$. This upper limit is below the one obtained from the astrophysical observations: $\Omega h^2_{100} \leq 10^{-5}$. If the sensitivity of testing the equivalence principle increases and the upper limit on $q_b^2$ is improved by one order of magnitude, the upper limit on $\Omega_{DW}$ that can be placed by the geophysical search becomes less stringent and comparable to the one obtained by the astrophysical observations (Figure \ref{fig:Constraint}).

I have used the recent upper bound on the effective viscosity of the liquid outer core for the estimation of the constraint. The determination of the effective viscosity is crucial for the accurate estimate of the constraint.

For simplicity, I have assumed massless dilatons. 
However, according to a scenario in string cosmology, a relic background of light dilatons could be a significant component of dark matter in the Universe \cite{Gasperini2003}. 
Unlike the massless background, the intensity of light mass-dilaton background is not constrained by the astrophysical observations and its dimensionless energy density could be as large as unity. This indicates that the signals could be larger by about five orders of magnitude than the one of the massless background; in spite of the weakness of the dilaton coupling, it might be within reach of experiments \cite{Gasperini2003}. The possibility of testing the cosmological scenario by the geophysical approach has to be studied further.  

\section{Conclusions}
In my previous work, I have proposed the geophysical search for dilatonic waves and presented preliminary upper limits on a massless dilatonic background by assuming a simple Earth model. In this work, I have discussed the upper limit further and presented the constraint on the dimensionless energy density of the massless dilaton background as a function of the strength of the dilaton coupling, $q_b^2$. The geophysical method is sensitive to the direct coupling of the dilaton fields to matter and the constraint obtained by this method depends on the strength of the dilaton coupling. 

Using the most stringent upper limit on $q_b^2$ by tests of the equivalence principle, we have obtained the upper limit on the dimensionless energy density of the massless dilaton background, $\Omega_{DW}h^2_{100} \leq 6 \times 10^{-7}$, at the frequency of $\sim$ 7 $\times$ 10$^{-5}$ Hz. This upper limit is about one-order of magnitude more stringent than the one from the astrophysical observations. If the magnitude of $q_b^2$ is experimentally found to be smaller than the current upper limit by one order of magnitude, the geophysical upper limit on $\Omega_{DW}h^2_{100}$ becomes less stringent and comparable to the one obtained from the astrophysical observations. 

I have discussed the massless dilaton background for simplicity. The constraint on the relic background of light dilatons, predicted in string cosmology, has to be studied in the future.

\ack
This work was funded by the Ministry of the Interior and National
Science Council of Taiwan.

\end{document}